\documentclass[12pt]{article}
\usepackage{epsfig}
\usepackage{psfig}
\begin{document}
\begin{titlepage}
\begin{center}

\vspace{5cm}

{\Large \bf Glueball-Induced Partonic Energy Loss in Quark-Gluon Plasma }
\vspace{0.50cm}\\Dong-Pil Min$^a$
\footnote{dpmin@snu.ac.kr} and  Nikolai
Kochelev$^{a,b}$\footnote{kochelev@theor.jinr.ru},
\vspace{0.50cm}\\
{(a) \it Department of Physics and Astronomy, Center for
Theoretical Physics, \\Seoul National University,
Seoul 151-747, Korea}\\
\vskip 1ex {(b) \it Bogoliubov Laboratory of Theoretical Physics,
Joint Institute for Nuclear Research, Dubna, Moscow region,
141980, Russia} \vskip 1ex

\end{center}
\vskip 0.5cm \centerline{\bf Abstract} We discuss the energy loss of
energetic parton jets in quark-gluon plasma above the
deconfinement temperature $T_c$ by
the interaction with scalar and pseudoscalar glueballs. It is shown
that the loss by this mechanism is quite important and may play the important role of the observed jet-quenching.
\vskip 0.3cm
\leftline{Pacs: 24.85.+p, 12.38.-t, 12.38.Mh, 12.39.Mk}
\leftline{Keywords: quarks, gluons, glueball, plasma,
non-perturbative QCD} \vspace{1cm}
\end{titlepage}
\setcounter{footnote}{0}


The investigation on jets in the relativistic heavy ion collisions (RHIC) provides us insights to understand the properties of quark-gluon plasma
(QGP)
\cite{STAR,PHENIX,BRAHMS,PHOBOS}. One of the important RHIC
discoveries is the jet quenching phenomenon coming from the partonic
energy loss in QGP. In the conventional approach to the jet
quenching the perturbative (pQCD) type of energy loss
is taken into account by the channels, elastic and radiative, of one-gluon exchange between
the jet and the massless gluons and quarks (see recent discussion in \cite{radiative}). However, the large quark-gluon rapidity
 density $dN_{qg}/dy\approx 2000$ which is needed to describe the RHIC jet quenching
 data within this approach, seems to be in contradiction with the
 restriction $dN_{qg}/dy\leq 1/4dS/dy\approx 1300$ coming from the
measured final entropy density $dS/dy\approx 5000$  \cite{muller}.
Furthermore, the lattice calculations show that even at very
high temperature
gluons and quarks still interact strongly in QGP
\cite{lattice1}.

Recently, it was suggested that the glueballs, the bound states of
gluons, can exist above deconfinement temperature and may play
an important role in the dynamic of strongly interacting QGP
\cite{vento,kochmin}. In particular, in \cite{kochmin}
it is suggested that a very light pseudoscalar glueball can
exist in QGP and might be responsible for the residual strong interaction
between gluons. The lattice results showing a change of sign of the gluon
condensate \cite{lattice1} and a small value of the topological
susceptibility \cite{lattice2} above $T_c$ can be explained in
the glueball picture as well.  Furthermore, one  expects that the
suppression of the mixing between  glueballs and quarkonium states in the QGP leads
 to a smaller width for former as compared to the  vacuum \cite{vento}. This property opens the
  possibility for a clear
separation of the glueball and the quark states in heavy ion collisions. Such separation
is rather difficult in other hadron reactions  due to  existence of strong
glueball-quarkonium mixing in the vacuum.

In this communication we report our study on the contribution of glueballs to the
energy loss by high energy partons propagated in QGP. We
will argue that significant partonic energy loss can result
from strong glueball-gluon coupling.


Our starting point is the effective pseudoscalar glueball-gluon
vertex employed in Refs.~\cite{kochmin, soni}
\begin{eqnarray}
{\cal L}_{Ggg}=\frac{1}{f_S}(\alpha_sG^a_{\mu\nu}{G}^a_{\mu\nu}S+
\xi\alpha_sG^a_{\mu\nu}\widetilde{G}^a_{\mu\nu}P), \label{lag}
\end{eqnarray}
where $G^a_{\mu\nu}$ is gluon field strength,
$\widetilde{G}^a_{\mu\nu}=\epsilon_{\mu\nu\alpha\beta}{G}^a_{\alpha\beta}/2$,
$ S $ and $P$ are scalar and pseudoscalar glueball fields,
respectively, $\xi\approx 1$, and
$f_S\approx 0.35$ GeV which is fixed by low energy theorem
\cite{kochmin}.
In Fig.1 the diagrams contributing to the energy loss in
high energy limit $s>> (-t,M^2_{S,P})$ are illustrated.

The elastic energy loss is given by Bjorken's formula
\cite{bjorken}
\begin{equation}
\frac{dE}{dx}(T)=\int d^3kn(k,T)[Flux factor]\int
dt\frac{d\sigma}{dt}\nu, \label{bj}
\end{equation}
where $\nu=E^\prime-E$ energy difference between fast incoming
and outcoming partons, $ [Flux factor]=(1-cos\theta)$, $\theta $
is the laboratory angle between the incident partons,
$d\sigma/dt$ is partonic cross section and $n(k,T)$ is density of
target parton in QGP at the temperature $T$.
\begin{figure}[h]
\centerline{\epsfig{file=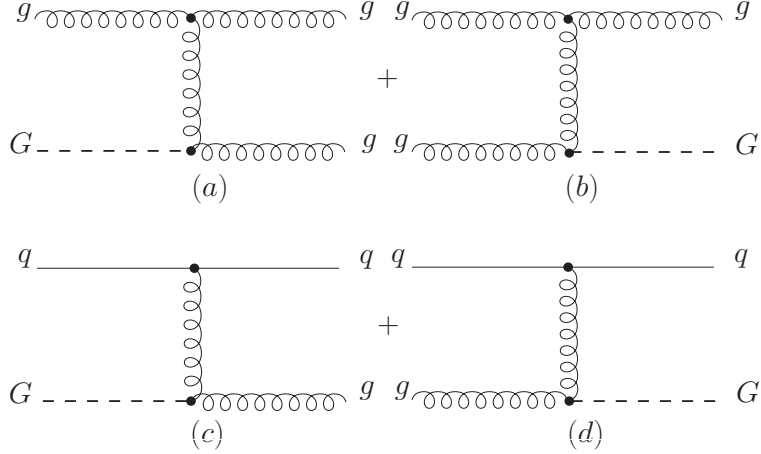,width=6cm,height=10cm,
angle=90}}\ \caption{ The diagrams contributed to a),b),d) gluon and
c),d) quark energy losses. The g (G) denotes gluon (glueball) and
q  the quark.}
\end{figure}
\begin{figure}
\begin{minipage}[c]{8cm}
\centerline{\epsfig{file=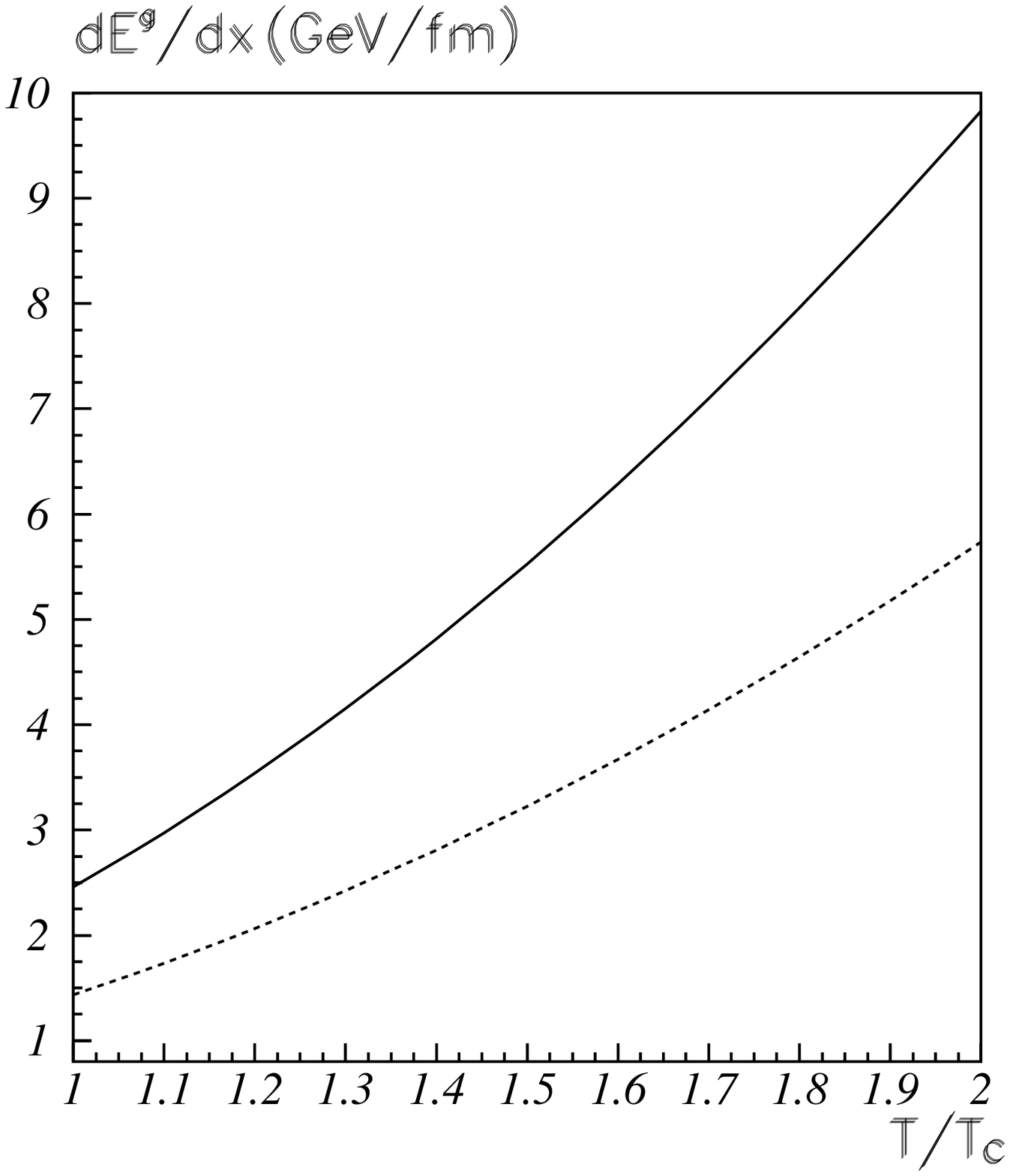,width=8cm, angle=0}}
\caption{
The temperature dependence of gluon energy loss. The solid
(dashed) line is glueball (pQCD) contribution.}
\end{minipage}
\hspace*{0.5cm}
\begin{minipage}[c]{8cm}
\centerline{\epsfig{file=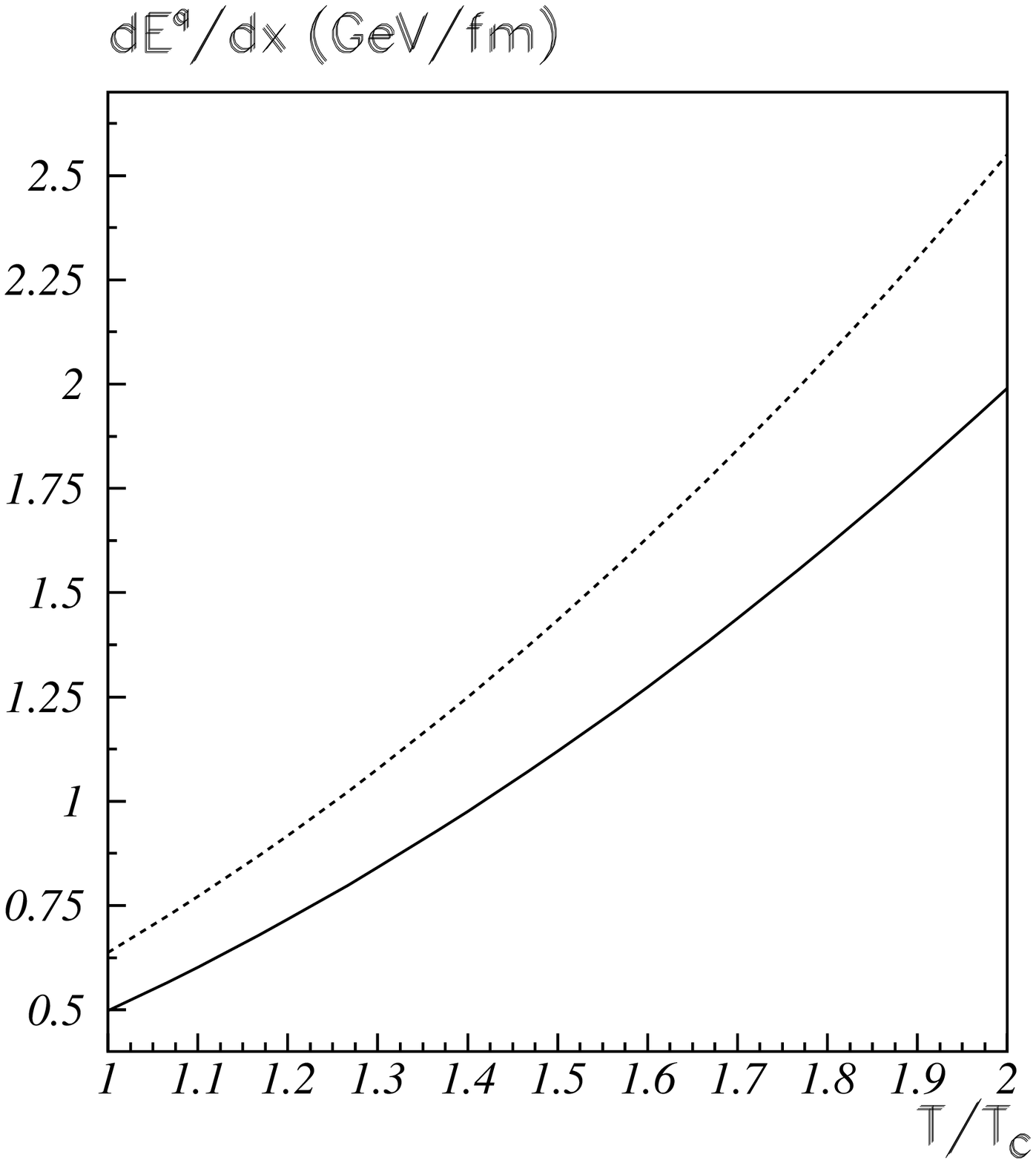,width=8cm, angle=0}}\
\vspace*{-0.5cm} \caption{ The temperature dependence of quark
energy loss. The notations are same as in Fig.2}
\end{minipage}
\end{figure}
The result of the
calculation of diagrams in Fig.1 is
\begin{eqnarray}
\frac{d\sigma_a}{dt}\approx\frac{15\alpha_s^3}{f_S^2|t|}F(t), \ \
\frac{d\sigma_b}{dt}\approx\frac{15\alpha_s^3}{4f_S^2|t|}F(t),\nonumber\\
\frac{d\sigma_c}{dt}\approx\frac{16\alpha_s^3}{3f_S^2|t|}F(t), \ \
\frac{d\sigma_d}{dt}\approx\frac{\alpha_s^3}{3f_S^2|t|}F(t),
\label{cross}
\end{eqnarray}
where $|t|=2k\nu(1-cos\theta)$, and the form factor in gluon-glueball
vertex reads \cite{kochmin}
\begin{equation}
F(t)=e^{-\Lambda^2|t|}, \label{cut}
\end{equation}
with $\Lambda\approx 0.6$ GeV$^{-1}$.
 In the high energy  limit   we  will neglect the small
effect coming from finite masses of the produced glueballs
 but we will  take into account the finite value of
 their masses in
the densities, Eq.\ref{dens}. Furthermore,
  our consideration here  is restricted
by calculation to the leading order in $\alpha_s$. So the possible thermal gluon
mass effects, $m_g\propto \alpha_sT$, are not considered. Therefore,
in the case of energetic parton it is enough to keep only
leading energy independent terms of the  partonic cross sections shown in
Eq.\ref{cross}.

The final result for energy loss for gluon and quark jets reads
\begin{eqnarray}
\frac{dE_{g}}{dx}(T)&=& \frac{15\alpha_s^3}{2f_S^2\Lambda^2}\int
\frac{d^3k}{k}[n_S(k,T)+n_P(k,T)+\frac{n_g(k,T)}{4}+\frac{n_q(k,T)}{45}]\nonumber\\
\frac{dE_{q}}{dx}(T)&=&\frac{8\alpha_s^3}{3f_S^2\Lambda^2}\int
\frac{d^3k}{k}[n_S(k,T)+n_P(k,T)+\frac{n_g(k,T)}{8}],
\label{loss2}
\end{eqnarray}
For estimation we will assume in QGP gluons, quarks and glueballs are
in thermodynamical equilibrium and will use the gas
approximation for gluon and glueball densities
\begin{equation}
n_{i}(k,T)=\frac{N_i}{(2\pi)^3(exp{(\sqrt{k^2+M^2_i}/T)}\pm
1)},
\label{dens}
\end{equation}
where the plus (minus) sign is for fermions (bosons) and  numbers of
degrees of freedom are $N_S=1$ for scalar  and $N_P=1$ for pseudoscalar glueballs,
respectively,
 $N_g=16$ for gluons and $N_q=12 $ for
number of light quark flavors $N_F=2$.
In our previous paper \cite{kochmin} it was argued that in the model with Lagrangian Eq.~\ref{lag}
the behaviour of the masses of pseudoscalar and scalar glueballs above $T_c$ is very different.
 Indeed, it was shown that the scalar glueball remain to be  massive, $M_S\approx 1.5$ GeV,
 but pseudoscalar
glueball is  very light, $M_P\approx 0$,  above deconfinement temperature.
Within  such approximation we obtain
\begin{equation}
\frac{dE_{g,q}}{dx}(T)\approx
C^G_{g,q}\frac{\alpha_s^3T^2}{f_S^2\Lambda^2},
\label{loss3}
\end{equation}
where  $C^G_g=79/24  $ is the coefficient for the glueball contribution to gluon energy loss and
 $C^G_q=2/3$ is the correspondent coefficient for quark energy loss. \footnote{ We neglect the
small contribution arising from the interaction of energetic
parton with massive scalar glueball in
QGP due to small value of its density $n_S<<n_P$.}

The numerical result presented in Figs.(2,3) of the temperature dependence of elastic
energy loss due to interaction with glueballs is to be compared with the recent re-analysis of perturbative QCD
elastic contribution in the range of temperatures $T_c<T<2T_c$, which is
accessible at RHIC experiments.
The pQCD elastic contribution is as following \cite{pQCD}
\begin{equation}
\frac{dE^{pQCD}_{g,q}}{dx}(T)=
C^{p}_{g,q}\frac{8\pi^2\alpha_sT^2}{b_0}(1+\frac{N_F}{6}),
\label{lossp}
\end{equation}
where $C^p_g=3/2, C^p_q=2/3$ are coefficients for the perturbative gluon and quark energy loss,
 respectively,
and $b_0=11/3N_c-2/3N_F$. We take
$T_c=170$ MeV for
$N_F=2$ \cite{lattice} and $\alpha_s\approx 0.6$ at
$T_c<T<2T_c$ \cite{latticealpha,alpha2} for the estimation of
energy loss in gluon-glueball plasma. It follows from Figs. 2,3
that glueball-induced energy loss is large for both gluon and
quark jets.  In particular, for the gluon jet such contribution is about of few GeV/fm and
 approximately twice larger than the perturbative elastic loss \cite{pQCD}.
In spite of the fact that
 for the quark jet the glueball contribution is  smaller than perturbative elastic loss, it can
 not be neglected in comparison with latter one. It is
evident that the origin of such large contribution is in
strong glueball-gluon coupling in Eq.\ref{lag}.
We should point
out that more than one half of contribution to the gluon energy
loss comes from interaction of gluon with the light
pseudoscalar glueball in QGP. Therefore, existence of such light
bound state of gluons above $T_c$ is crucial for the
understanding of the large observed partonic energy loss in QGP.

In summary, we made the estimation of the energy loss induced by
interaction of an energetic parton,  which was produced  in the  hard scattering of two heavy ion's
partons, with glueballs in hot quark-gluon
plasma. It is shown that such contribution leads to a
significant energy loss. We conclude that not only pQCD type of
energy loss but also glueball-induced loss, arising from existence
of scalar and pseudoscalar glueballs in QGP, are important for the
understanding of the RHIC results such as the jet quenching.
 We should emphasize
that the main goal of our paper is  to show the significance  of the
glueball-induced energy loss and we left the detailed comparison with experiment,
which should
include also the consideration of the effects of both  elastic and radiative pQCD losses,
 for a forthcoming
publication.\\


{\bf Acknowledgments}\\

We would like to thank V.Vento for useful discussions.
This work was supported by Brain Pool program of Korea Research
Foundation through KOFST, grant 042T-1-1. NK is very grateful to
the School of Physics and Astronomy of Seoul National University
for their warm hospitality during this work.

\end{document}